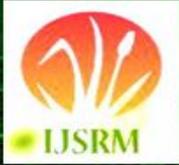
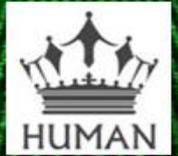

# On the Origin of the Lorentz Transformation

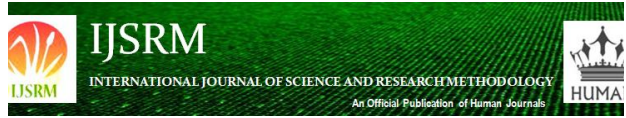

**W. Engelhardt**

*Retired from: Max-Planck-Institut für Plasmaphysik, D-85741 Garching, Germany*

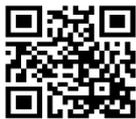
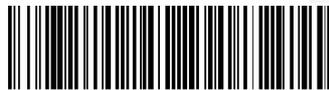

www.ijsrm.humanjournals.com

**Keywords:** Special relativity, Voigt, Lorentz, Poincaré, Einstein

**ABSTRACT**

The Lorentz Transformation, which is considered as constitutive for the Special Relativity Theory, was invented by Voigt in 1887, adopted by Lorentz in 1904, and baptized by Poincaré in 1906. Einstein probably picked it up from Voigt directly.

## 1 INTRODUCTION:

**Maxwell's wave equation**

Maxwell's ether theory of light was developed from his first order equations and resulted in a homogeneous wave equation for the vector potential:

$$c^2 \Delta \vec{A} = \frac{\partial^2 \vec{A}}{\partial t^2}$$

The electromagnetic field could be calculated by differentiating this potential with respect to time and space: $\vec{E} = -\partial \vec{A}/\partial t$, $\vec{B} = \mathrm{rot}\, \vec{A}$. For a wave polarized in $y$-direction, e.g., and travelling in $x$-direction one may write for the $y$-component of $\vec{A}$:

$$c^2 \frac{\partial^2 A}{\partial x^2} = \frac{\partial^2 A}{\partial t^2} \qquad (1)$$

Formally this equation is the same as a wave equation for sound, e.g.

$$c_s^2 \frac{\partial^2 p}{\partial x^2} = \frac{\partial^2 p}{\partial t^2} \qquad (2)$$

Where $c_s$ is the sound velocity in air and $p$ is the pressure perturbation. Maxwell took the propagation velocity in (1) as the velocity of light with respect to the ether that was perceived by him as a hypothetical medium in which light propagates like sound in air.

From the formal similarity of (1) and (2) follow similar solutions, e.g. plane waves:

$$A = A_0 \cos(k\,x - \omega t) \qquad (3)$$

At a fixed time $t$ the wave has an oscillatory behaviour in space with wavelength $\lambda = 2\pi/k$, and at a fixed point $x$ the amplitude $A$ of the wave oscillates with frequency $\nu = \omega/2\pi$. A point of fixed phase, e.g. $k\,x - \omega t = 0$, travels with phase velocity

$$\frac{x}{t} = \frac{\omega}{k} = c \qquad (4)$$

This result is easily obtained by inserting (3) into (1).

If an observer travels along the *x*-axis, i.e. parallel to the $\vec{k}$ -vector, with velocity $v$, we have the obvious Galilei connection between his coordinates ($x´,t´$) and the wave coordinates:

$$x´= x - vt, \quad t´= t \qquad (5)$$

Substituting this into (3) one obtains again a plane wave in the moving system:

$$A = A_0 \cos(k(x´+vt´) - \omega t´) = A_0 \cos(k x´-(\omega-kv)t´) \qquad (6)$$

The phase velocity as observed in that system is:

$$\frac{x´}{t´} = \frac{\omega}{k} - v = c - v \qquad (7)$$

and the frequency becomes

$$\omega´ = \omega - kv = \omega(1 - v/c) \qquad (8)$$

according to (6) and (4). This equation reflects the Doppler Effect which is well known from observations both in sound and in light.

Due to the change of the phase velocity (7) it is clear that equation (1) must change its form, when transformed into a moving system. Using the chain rule for differentiation we have from eq. (5):

$$\frac{\partial}{\partial x} = \frac{\partial}{\partial x´}\frac{\partial x´}{\partial x} + \frac{\partial}{\partial t´}\frac{\partial t´}{\partial x} = \frac{\partial}{\partial x´}, \quad \frac{\partial}{\partial t} = \frac{\partial}{\partial x´}\frac{\partial x´}{\partial t} + \frac{\partial}{\partial t´}\frac{\partial t´}{\partial t} = -v\frac{\partial}{\partial x´} + \frac{\partial}{\partial t´}$$
$$\frac{\partial^2}{\partial x^2} = \frac{\partial^2}{\partial x´^2}, \quad \frac{\partial^2}{\partial t^2} = v^2\frac{\partial^2}{\partial x´^2} - 2v\frac{\partial^2}{\partial x´\partial t´} + \frac{\partial^2}{\partial t´^2} \qquad (9)$$

Substituting these expressions into (1) yields the wave equation in the moving system:

$$c^2\frac{\partial^2 A}{\partial x´^2} = v^2\frac{\partial^2 A}{\partial x´^2} - 2v\frac{\partial^2 A}{\partial x´\partial t´} + \frac{\partial^2 A}{\partial t´^2} \qquad (10)$$

Inserting the plane wave ansatz $\cos(k x´-\omega´ t´)$, one obtains the dispersion relation

$$(c^2 - v^2)k^2 = 2vk\omega´ + \omega´^2 \qquad (11)$$

in agreement with (4) and (8).

## 2 Voigt's Theory of the Doppler Effect

Although the theory developed so far describes very well the Doppler Effect both for sound and light, Woldemar Voigt published an alternative theory for elastic media in 1897 which he called *On Doppler´s Principle*[i]. He insisted that equation (1) should maintain its form upon transformation into a moving system, namely equation (10) should read:

$$c^2 \frac{\partial^2 A}{\partial x'^2} = \frac{\partial^2 A}{\partial t'^2} \qquad (12)$$

This amounts to requiring that the phase velocity of a wave is independent of the motion of the observer. It is unclear how Voigt arrived at this curious idea. He simply commented eq. (12) with the cryptic remark "*da ja sein muss*", "as it must be". We cannot exclude that he was mistaken by a misconception of the nature of propagating waves. In any case, in order to achieve his goal he transformed time and made it a linear function of space. Instead of (5) he wrote

$$x' = x - vt, \quad t' = t - xv/c^2 \qquad (13)$$

Now the phase velocity in the moving system becomes with (13):

$$c' = \frac{x'}{t'} = \frac{x - vt}{t - xv/c^2} \qquad (14)$$

Inserting $x = ct$ from (4) one obtains instead of (7):

$$c' = \frac{ct - vt}{t - tv/c} = c \qquad (15)$$

Voigt did apparently not realize that the independence of the phase velocity from the motion of the observer is in blatant contradiction to the observations in sound for which his theory was supposed to hold as well. Obviously, it is inconceivable that time transforms in compliance with all the different wave velocities occurring in elastic media.

For reasons we do not know either, Voigt's transformation appeared attractive to Lorentz in his pursuit to develop an electrodynamics for moving media. In his paper of 1904[ii] he simply says: "I take as new independent variables…" and introduces subsequently Voigt's transformation (13) without mentioning his name. In a monograph of 1913[iii], however, which contained early papers related to relativity by Lorentz, Einstein, and Minkowski, he added a footnote where he acknowledged the equivalence of his transformation to that of Voigt's from 1887.

In his paper of 1906[iv] Poincaré referred not to Voigt's, but to Lorentz's work and gave the transformation (13) Lorentz's name. It appears that he was intrigued by the postulated constancy of the light velocity for all observers regardless of their motion. This seemed to be in accordance with his general principle of relativity that would make it impossible to distinguish different inertial states of motion by electromagnetic measurements. Poincaré also recognized that the linear relationship (13) could be multiplied by a constant factor which cancels in (14), thus keeping the velocity of light still constant. The final form of the "Lorentz Transformation" took then the symmetrical shape:

$$x' = \gamma(x - vt), \quad y' = y, \quad z' = z, \quad t' = \gamma(t - xv/c^2), \quad \gamma = 1/\sqrt{1 - v^2/c^2} \tag{16}$$

Which is form-invariant to the inverse transformation choosing $v \rightarrow -v$:

$$x = \gamma(x' + vt'), \quad y = y', \quad z = z', \quad t = \gamma(t' + x'v/c^2) \tag{17}$$

It remains to demonstrate that application of (16) transforms indeed eq. (1) into eq. (12). Instead of (9) we have from (16):

$$\frac{\partial}{\partial x} = \frac{\partial}{\partial x'}\frac{\partial x'}{\partial x} + \frac{\partial}{\partial t'}\frac{\partial t'}{\partial x} = \gamma\left(\frac{\partial}{\partial x'} - \frac{v}{c^2}\frac{\partial}{\partial t'}\right), \quad \frac{\partial}{\partial t} = \frac{\partial}{\partial x'}\frac{\partial x'}{\partial t} + \frac{\partial}{\partial t'}\frac{\partial t'}{\partial t} = \gamma\left(-v\frac{\partial}{\partial x'} + \frac{\partial}{\partial t'}\right)$$

$$\frac{\partial^2}{\partial x^2} = \gamma^2\left(\frac{\partial^2}{\partial x'^2} - \frac{2v}{c^2}\frac{\partial^2}{\partial x'\partial t'} + \frac{v^2}{c^4}\frac{\partial^2}{\partial t'^2}\right), \quad \frac{\partial^2}{\partial t^2} = \gamma^2\left(v^2\frac{\partial^2}{\partial x'^2} - 2v\frac{\partial^2}{\partial x'\partial t'} + \frac{\partial^2}{\partial t'^2}\right) \tag{18}$$

If one substitutes these expressions into (1), one obtains:

$$c^2 \frac{\partial^2 A}{\partial x'^2}\gamma^2\left(1 - v^2/c^2\right) = \frac{\partial^2 A}{\partial t'^2}\gamma^2\left(1 - v^2/c^2\right) \tag{19}$$

where the $\gamma$-factor cancels.

## 3 Einstein's *procedere*

Einstein like Poincaré was convinced of the validity of the generalized relativity principle. In his famous paper of 1905[v] he raised this conjecture to a postulate and claimed that he could derive the Lorentz Transformation by using a second postulate, namely "that light is always propagated in empty space with a definite velocity *c* which is independent of the state of motion of the emitting body". A careful analysis of §3 of his paper shows, however, that he worked on the same assumption as Voigt, namely that the velocity of light is constant for any moving observer. This is reflected in the little relationship on page 898

$$\tfrac{1}{2}(\tau_0 + \tau_2) = \tau_1$$

which Einstein introduced with nearly the same wording as Voigt: *"so muss dann sein"*, "thus there must be". Here $\tau_0$ denotes the time when a light signal is emitted from the origin of a moving coordinate system to some point on the *x*-axis where it arrives at $\tau_1$, and returns to the origin at $\tau_2$. Obviously, this inconspicuous relationship implies $c = $ const in the moving system as a precondition for the derivation. The technical term for taking the conclusion to be proved into the premise is *petitio principii*. The first postulate, namely his "principle of relativity", is not used in Einstein's derivation of the Lorentz Transformation.

It is likely that Einstein, who did not quote anybody in Ref. 5, found his transformation directly in Voigt's paper of 1897. In the monograph of 1913 (Ref. 3) he assures us that he was not aware of Lorentz's paper of 1904. This is credible, as he did not use Lorentz's nomenclature $(x, y, z, t) \leftrightarrow (x', y', z', t')$, but he adopted Voigt's: $(x, y, z, t) \leftrightarrow (\xi, \eta, \zeta, \tau)$. He also added a footnote where he stated that the "Lorentz Transformation" is much easier to be derived using Voigt's postulate, but – in contrast to Lorentz – he avoided actually to mention his name. In the Appendix we quote Einstein's "Simple derivation of the Lorentz Transformation" of 1917.

It is instructive to recognize that Voigt's absurd claim of the constancy of a wave's phase velocity makes as little sense in electrodynamics as it does in acoustics. The linear Doppler Effect as described in (8) is normally thought to be due to different velocities of an observer with respect to the wave crests. For a wave we have the well known relationship:

$$c = \lambda \nu \qquad (20)$$

When we measure an increased frequency $\nu'$, more wave crests hit the detector per unit time, because the detector has an increased relative velocity $c+v$ with respect to the wave fronts. If one claims, however, that the relative velocity between detector and wave front is always $c$, one must conclude from (20) and (8) that the wavelength transforms linearly with the velocity:

$$\frac{\lambda'}{\lambda} = \frac{\nu}{\nu'} = \frac{c}{c \pm v} \qquad (21)$$

This relationship is not often discussed in textbooks on SRT, probably because of its absurd consequences. If the length of light waves would transform linearly with velocity, but the distance between mirrors in a resonator transforms quadratically, as the Lorentz contraction assumes, a laser would be extinguished, simply because an observer passes by at high velocity, thereby destroying the resonance condition which requires an integral number of modes in the resonator. Nobody believes this, but one uses the linear Doppler Effect to measure the velocity of the earth relative to the wave crests of the cosmic background radiation and finds 370 km/s versus τ Leonis in obvious violation of the generalized relativity principle.

Concluding this little essay I would like to communicate an interesting aperçu. Einstein kept denying that the Michelson-Morley experiment had any influence on the development of his theory [vi]. Probably he had read a paper with the title *Theory of light for moving media* by W. Voigt[vii] which appeared two months later than Ref. 1. On page 233 Voigt analyzes the Michelson experiment and comes to the conclusion: „Hieraus folgt, daß die Beobachtungsresultate bei der Anordnung des Experimentes, wie sie Herr M i c h e l s o n gewählt hat, von der Translation völlig unabhängig sind, daß also Herr Michelson die negativen Resultate, die er factisch erhalten hat, erhalten m u ß t e, gleichviel, ob sich der Aether mit der Erde bewegt oder nicht." [viii] If Voigt was right, Einstein could not possibly take Michelson´s result as experimental evidence for the correctness of his theory. It is likely that he was aware of Voigt's paper and of his judgement. And if this was the case, he had certainly also read Ref. 1 which offered him a transformation he could exploit for his purpose.

**Appendix**

In Einstein´s book of 1917: *Über die spezielle und die allgemeine Relativitätstheorie* (http://www.ideayayinevi.com/metinler/relativitetstheorie/oggk00.htm) one finds an Appendix:

**Einfache Ableitung der LORENTZ-Transformation**.

To start with, Einstein introduces two equations

$$x - ct = 0 \qquad (1)$$

$$x' - ct' = 0 \qquad (2)$$

Subsequently a quantity $\lambda$ is defined by a third equation:

$$x - ct = \lambda (x' - ct') \qquad (3)$$

If one substitutes (1) and (2) into (3) one obtains:

$$0 = \lambda\, 0$$

From this and a similar equation

$$0 = \mu\, 0 \qquad (4)$$

Albert Einstein derives the Lorentz Transformation. In the Preface he writes:

*Das vorliegende Büchlein soll solchen eine möglichst exakte Einsicht in die Relativitätstheorie vermitteln, die sich vom allgemein wissenschaftlichen, philosophischen Standpunkt für die Theorie interessieren, ohne den mathematischen Apparat der theoretischen Physik zu beherrschen. Die Lektüre setzt etwa Maturitätsbildung und — trotz der Kürze des Büchleins — ziemlich viel Geduld und Willenskraft beim Leser voraus.*

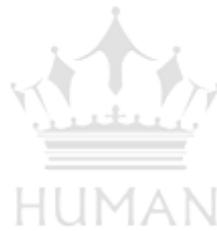